\documentclass[showpacs,preprint]{revtex4-1}
\usepackage{amssymb}
\usepackage{amsmath}
\usepackage{graphicx}

\begin{document}

\title{Nucleon distribution in nuclei beyond $\beta $-stability line}
\author{V.M. Kolomietz, S.V. Lukyanov and A.I. Sanzhur}

\begin{abstract}
The radii of nucleon distribution, bulk density, and neutron skin in nuclei
beyond the $\beta $-stability line are studied within the direct variational
method. We evaluate the partial equation of state of finite nuclei and
demonstrate that the bulk density decreases beyond the beta stability line.
We show that the growth of the neutron skin in unstable nuclei does not obey the
saturation condition because of the polarization effect. The value of the
neutron-skin thickness $\Delta r_{np}=\sqrt{\left\langle
r_{n}^{2}\right\rangle }-\sqrt{\left\langle r_{p}^{2}\right\rangle}$ is
caused by the different radii (skin effect) and only slightly by the
different shapes (halo effect) of neutron and proton distributions. The
relative contribution of both effects depends on the competition between the 
symmetry energy, and the spin-orbit and Coulomb interactions. The
calculations of the isovector shift of the nuclear radius $\Delta r_{np}$
show its primarily linear dependence on the asymmetry parameter $X=(N-Z)/A$.
\end{abstract}

\maketitle

\affiliation{Institute for Nuclear Research, 03680 Kiev, Ukraine}

\section{Introduction}

Our knowledge about the properties of neutron excess in heavy nuclei and its
relation to the neutron-rich nuclear matter and the isotopic symmetry
energy is still strongly limited. In heavy stable nuclei, the average
changes in binding energy $E$ and nuclear radius $R$ with nucleon content
obey the saturation properties. The volume part $E_{\mathrm{vol}}$ of
binding energy and the nuclear volume itself are proportional to the
particle number $A$ with $E_{\mathrm{vol}}$ $=-\ b_{V}A$ and $R=r_{0}A^{1/3}$%
, where $b_{V}>0$ and $r_{0}$ are the constants. Both values of $b_{V}$ and $%
r_{0}$ depend, however, on the isotopic asymmetry parameter $X=(N-Z)/(N+Z)$.
This is because of the difference in saturation bulk density, $\rho _{0}\sim
r_{0}^{-3}$, of nuclei with different values of $X$. The saturation density $%
\rho _{0}$ is smaller beyond the beta-stability line for neutron-rich nuclei
where more neutrons are pushed off to form the ``neutron coating". 
One can expect then that the growth of the neutron skin in neutron-rich nuclei violates
the saturation property $R\sim A^{1/3}$ for the nuclear radius providing a
relative shift of both neutron and proton distributions \cite{meto02}. The
main characteristic of the neutron skin is the neutron-skin thickness  
$\Delta r_{np}=\sqrt{\left\langle r_{n}^{2}\right\rangle }-
\sqrt{\left\langle r_{p}^{2}\right\rangle }$, where $\sqrt{\left\langle
r_{n}^{2}\right\rangle }$ and $\sqrt{\left\langle r_{p}^{2}\right\rangle }$
are the neutron and proton root mean square (rms) radii , respectively. The
value of $\Delta r_{np}$ can be caused by the different radii (skin effect)
and the different shapes (halo effect) of neutron and proton distributions;
see also Refs. \cite{midolanare00,trja01,krak04,waviroce10,ceroviwa10}. The
relative contribution of both effects depends on the competition between
symmetry energy, spin-orbit and Coulomb interactions \cite%
{niscnabepe11,gansamo11}.

In the present paper we study a deviation of nucleon distribution from the
saturation behavior in neutron-rich nuclei. We consider the influence of the
spin-orbit and Coulomb forces on the neutron, $\sqrt{\left\langle
r_{n}^{2}\right\rangle }$, and proton, $\sqrt{\left\langle
r_{p}^{2}\right\rangle }$, \textrm{rms} radii as well as the relation of the
shift $\Delta r_{np}$ to the surface symmetry energy. We study also the
related problems of the nucleon redistribution within the surface region
(nuclear periphery), in particular, the neutron coating and the neutron excess
for the nuclei far away from the $\beta$-stability line.

We combine the extended Thomas-Fermi (\textrm{ETF}) approximation which
takes into consideration the corrections up to the order of $\hbar^{2}$ and the
direct variational method assuming that the proton and neutron distributions
are sharp enough, i.e., that the corresponding densities $\rho _{p}(\mathbf{r%
})$ and $\rho _{n}(\mathbf{r})$ fall from their bulk values to zero in a
thin region around the surface. In our consideration, the thin-skinned
densities $\rho _{p}(\mathbf{r})$ and $\rho _{n}(\mathbf{r})$ are generated
by the profile functions which are eliminated by the requirement that the
energy of the nucleus should be stationary with respect to variations of
these profiles. Note that the use of the direct variational method and the
trial profile function for the particle density allows us to derive the
equation of state (the dependence of the pressure on the bulk density) in
the case of the finite diffuse layer of the particle distribution in finite
nuclei.

This paper is organized as follows. In Sec. \textrm{II} we formulate the
direct variational principle for the density profile function within the
extended Thomas-Fermi approximation. Using the leptodermous assumption, we
obtain $A^{1/3}$ expansion for the symmetry energy and the related values.
The results of numerical calculations are presented in Sec. \textrm{III}.
We conclude and summarize in Sec. \textrm{IV}.

\section{Direct variational approach}

We will use the extended Thomas-Fermi approximation which is one of the
practical realizations of the general Hohenberg-Kohn theorem \cite{hoko64} on the
unique functional relation between the ground-state energy and the local
density of particles for any fermion system. The key point of the \textrm{ETF} 
is that the total kinetic energy of the many-body fermion system is given
by the semiclassical expression \cite{kirz67,brguha85,book} as follows:
\begin{equation}
E_{\mathrm{kin}}\{\rho _{n},\rho _{p}\}\equiv E_{\mathrm{kin}}\{\rho _{q},%
\mathbf{\nabla }\rho _{q}\}=\int d\mathbf{r}\,\,\epsilon _{\mathrm{kin}%
}[\rho _{n}(\mathbf{r}),\rho _{p}(\mathbf{r})],  \label{ekin}
\end{equation}%
where $\epsilon _{\mathrm{kin}}[\rho _{n},\rho _{p}]=\epsilon _{\mathrm{kin,}%
n}[\rho _{n}]+\epsilon _{\mathrm{kin,}p}[\rho _{p}]$, and

\begin{equation}
\epsilon _{\mathrm{kin,}q}[\rho _{q}]={\frac{\hbar ^{2}}{2m}}\left[ {\frac{3%
}{5}}\,(3\,\pi ^{2})^{2/3}\,\rho _{q}^{5/3}+\beta {\frac{(\mathbf{\nabla }%
\rho _{q})^{2}}{\rho _{q}}}+{\frac{1}{3}}\,\nabla ^{2}\rho _{q}\right] \,.
\label{ekin2}
\end{equation}%
Here $\rho _{q}$ is the nucleon density with $q=n$\ for neutron and $q=p$
for proton. The semiclassical consideration gives the value of parameter $%
\beta $ in Eq. (\ref{ekin2}) $\beta =1/36$ \cite{kirz67,brguha85}. We point
out that in the asymptotic limit $r\rightarrow \infty $, the semiclassical
particle density $\rho _{q}$ with $\beta =1/36$\ goes significantly faster
to zero than the one from the quantum-mechanical calculation. An asymptotic
solution for the particle density $\rho _{\mathrm{ETF},q}$\textbf{$(\mathbf{r%
})$} within the semiclassical extended Thomas-Fermi approximation in the
limit $r\rightarrow \infty $ has the following form \cite{brguha85}:

\begin{equation}
\left. \rho_{\mathrm{ETF},q}(\mathbf{r})\right\vert _{r\rightarrow \infty
}\sim {\frac{1}{r^{2}}}\,\exp \left[ {-\sqrt{-{\frac{2m}{\hbar ^{2}}}\,{%
\frac{\lambda _{q}}{\beta }}}\,r}\right] ,  \label{rho1x}
\end{equation}%
where $\lambda _{q}$ is the chemical potential which is negative for a bound
Fermi system. A quantum-mechanical wave function whose bound $s$ orbital has
energy $\epsilon $, gives the partial contribution to the asymptotic
particle density, which is

\begin{equation}
\left. \rho _{\mathrm{part}}(\mathbf{r})\right\vert _{r\rightarrow \infty
}\sim {\frac{1}{r^{2}}}\,\exp \left[ {-2\,\sqrt{-{\frac{2\,m}{\hbar ^{2}}}%
\,\epsilon }\,r}\right] .  \label{rho2x}
\end{equation}

Thus, the semiclassical particle density $\rho _{\mathrm{ETF},q}(\mathbf{r})$
of Eq. (\ref{rho1x}) for $\beta =1/36$ goes faster to zero than the
quantum-mechanical one (\ref{rho2x}). To overcome this defect of the
extended Thomas-Fermi approximation the value $\beta $ can be considered as
an adjustable parameter. We will apply both the semiclassical value $\beta
=1/36$ and the phenomenological one $\beta =1/9$ which is consistent with
the quantum-mechanical asymptotic behavior given by Eq. (\ref{rho2x}).

We will follow the concept of the effective nucleon-nucleon interaction
using the Skyrme-type force. The functional of the total energy of charged
nucleus is given by 
\begin{equation}
E_{\mathrm{tot}}\{\rho _{q},\mathbf{\nabla }\rho _{q}\}=E_{\mathrm{kin}%
}\{\rho _{q},\mathbf{\nabla }\rho _{q}\}+E_{\mathrm{pot}}\{\rho _{q},\mathbf{%
\nabla }\rho _{q}\}+E_{\mathrm{C}}\{\rho _{p}\},  \label{etot}
\end{equation}%
where $E_{\mathrm{pot}}\{\rho _{q},\mathbf{\nabla }\rho _{q}\}$ is the
potential energy of $NN$ interaction
\begin{equation}
E_{\mathrm{pot}}\{\rho _{q},\mathbf{\nabla }\rho _{q}\}=\int d\mathbf{r}%
\,\,\epsilon _{\mathrm{pot}}[\rho _{n}(\mathbf{r}),\rho _{p}(\mathbf{r})],
\label{epot}
\end{equation}%
$\epsilon _{\mathrm{pot}}[\rho _{n}(\mathbf{r}),\rho _{p}(\mathbf{r})]$ is
the density of the potential energy of the nucleon-nucleon interaction and $%
E_{\mathrm{C}}\{\rho _{p}\}$ is the Coulomb energy. In our consideration,
the potential energy $E_{\mathrm{pot}}\{\rho _{q},\mathbf{\nabla }\rho
_{q}\} $ includes the energy of the spin-orbit interaction also.

Following the direct variational method, we have to choose the trial
function for $\rho _{q}(\mathbf{r})$. We will assume a power of the Fermi
function for $\rho _{q}(\mathbf{r})$ as 
\begin{equation}
\rho _{q}(\mathbf{r})=\rho _{0,q}\left[ 1+\exp \left( \frac{r-R_{q}}{a_{q}}%
\right) \right] ^{-\eta },  \label{rhoq}
\end{equation}%
where $\rho _{0,q}$, $R_{q}$, $a_{q}$, and $\eta$ are the unknown
variational parameters. Considering the asymmetric nuclei with $X=(N-Z)/A\ll
1$, we will introduce the isotopic particle densities, namely the total
density $\rho _{+}=\rho _{n}+\rho _{p}$ and the neutron excess density 
$\rho_{-}=\rho _{n}-\rho _{p}$ with $\rho _{-}\ll \rho _{+}$. Assuming a small
deviation of the isoscalar bulk density $\rho _{0,+}=\rho_{0,n}+\rho _{0,p}$, 
the radii $R_{q}$, and the diffuseness parameters $a_{q}$ with respect to
the corresponding average values of $\rho_{0}$, $R$, and $a$, we introduce
the density profile functions $\rho _{+}(r)$ and $\rho _{-}(r)$ to be given
by

\begin{equation}
\rho _{+}(r)=\rho _{0}\ f(r)-\frac{1}{2}\rho _{1}\frac{df(r)}{dr}\left[
\Delta _{R}+\frac{r-R}{a}\Delta _{a}\right] ,\quad \rho _{-}(r)=\rho _{1}\
f(r)-\frac{1}{2}\rho _{0}\frac{df(r)}{dr}\left[ \Delta _{R}+\frac{r-R}{a}%
\Delta _{a}\right] .  \label{prof1}
\end{equation}%
Here, 
\begin{equation}
f(r)=\left[ 1+\exp \left( \frac{r-R}{a}\right) \right] ^{-\eta },  \label{fr}
\end{equation}%
the values $\rho _{0}$ and $\rho _{1}$ are related to the bulk density, $R$
is the nuclear radius, $a$ is the diffuseness parameter, and $\Delta_{R}=R_{n}-R_{p}$ 
and $\Delta _{a}=$ $a_{n}-a_{p}$ are the parameters of the
neutron skin. The profile functions $\rho _{+}(r)$ and $\rho_{-}(r)$ have
to obey the condition that the number of neutrons and protons is conserved.
For the ground state of the nucleus, the unknown parameters $\rho _{0},$ $\rho
_{1},\ R,\ a,\ \Delta _{R}$, $\Delta_{a}$, and $\eta $ and the total energy $%
E_{\mathrm{tot}}$ itself can be derived from the variational principle 
\begin{equation}
\delta (E-\lambda _{n}N-\lambda _{p}Z)=0,  \label{var1}
\end{equation}%
where the variation with respect to all possible small changes of $\rho
_{0}, $ $\rho _{1},\ R,\ a,\ \Delta _{R}$, $\Delta _{a}$, and $\eta $ is
assumed. The Lagrange multipliers $\lambda _{n}$ and $\lambda _{p}$ are the
chemical potentials of the neutrons and the protons, respectively, and both of
them are fixed by the condition that the number of particles is conserved.

We point out that the bare nucleon mass $m$, but not the effective one $%
m^{\ast }$, enters the kinetic energy density $\epsilon _{\mathrm{kin,}%
q}[\rho _{q}]$ in Eq. (\ref{ekin2}) and accompanies the direct variational
procedure of Eq. (\ref{var1}). This is due to the fact that the expression (%
\ref{ekin2}) is directly\ obtained as a result of a Wigner transform to the
quantum-mechanical kinetic energy of many-body systems where only the bare mass 
$m$ is available; see, e.g., Ref. \cite{kirz67}. \ Usually, the effective
mass $m^{\ast }$ appears in the self-consistent Euler equations for the
particle density $\rho _{q}$ within the Thomas-Fermi approach (or for the
single-particle wave functions in the case of Hartree-Fock theory) after the
implementation of the corresponding variational procedure. The effective
mass appears there because the part of the self-consistent mean field, which is
caused by the non-local interparticle interaction, is associated with the
single-particle kinetic energy. In contrast, in our direct variational
method we do not use the Euler equations for the particle density. Thereby,
the effective mass $m^{\ast}$ cannot be included in the kinetic energy
density $\epsilon _{\mathrm{kin,}q}[\rho _{q}]$ of Eq. (\ref{ekin2}) to
avoid a twofold account of contributions from the $t_{1}-$ and $t_{2}-$
components of Skyrme forces which enter already the potential energy density 
$\epsilon _{\mathrm{pot}}[\rho _{q}(r)]$ in Eq. (\ref{epot}). This argument
is also applied to the spin-orbit interaction in a finite system where the
spin-orbit contribution to the energy functional $E_{\mathrm{pot}}\{\rho
_{q},\mathbf{\nabla }\rho _{q}\}$ of Eq. (\ref{epot}) is involved in the
direct variational procedure as well.

We will also assume that the leptodermous condition $a/R\ll 1$ is fulfilled.
The total energy (equation of state) (\ref{etot}) takes then the following
form \cite{kosa08}:
\begin{equation}
E_{\mathrm{tot}}(\rho _{0},X)/A=e_{0}(\rho _{0})\ +b_{\mathrm{S}}(\rho
_{0})A^{-1/3}+\left[ b_{\mathrm{V,sym}}(\rho _{0})+b_{\mathrm{S,sym}}(\rho
_{0})\ A^{-1/3}\right] X^{2}+E_{C}(\rho _{0},X)/A,  \label{e1}
\end{equation}%
where $e_{0}(\rho _{0})$ is the energy per nucleon of symmetric nuclear
matter, $b_{\mathrm{S}}(\rho _{0})$ is the surface energy coefficient, $b_{%
\mathrm{V,sym}}(\rho _{0})$ is the volume part of symmetry energy
coefficient, $b_{\mathrm{S,sym}}(\rho _{0})$ is the surface part of the
symmetry energy coefficient, and $E_{C}(\rho _{0},X)$ is the total Coulomb
energy 
\begin{equation}
E_{C}(\rho _{0},X)=\alpha _{C}(\rho _{0})\left( 1-X\right)
^{2}A^{5/3}+O(A^{4/3}),\quad \alpha _{C}(\rho _{0})=\frac{3}{20}e^{2}\left( 
\frac{4\pi \rho _{0}}{3}\right) ^{1/3}.  \label{eclmb}
\end{equation}%
The equation of state in the form of Eq. (\ref{e1}) implies that
the total energy per particle $E_{\mathrm{tot}}(\rho _{0},X)/A$ is minimized
with respect to the independent parameters $a,\ \Delta _{R}$, $\Delta _{a}$,
and $\eta $ for arbitrary values of $\rho _{0}$ and $X$.

The structure of the equation of state (\textrm{EOS)} given by Eq. (\ref{e1}) 
is similar to the semiempirical mass formula which describes the average
changes in nuclear binding energy with the mass number. However, in contrast
to the mass formula, the bulk density $\rho _{0}$ and the asymmetry
parameter are not necessarily at equilibrium. The symmetry term $\sim X^{2}$
includes both the volume, $b_{\mathrm{V,sym}}(\rho _{0})$, and the surface, $%
b_{\mathrm{S,sym}}(\rho _{0})$, contributions. The surface symmetry term $b_{%
\mathrm{S,sym}}(\rho _{0})\ A^{-1/3}X^{2}$ appears in the advanced mass
formula by Myers and Swiatecki \cite{mysw69,mysw74} and it is currently
employed in the description of surface properties and isovector excitations
in finite nuclei; see, e.g., Refs. \cite{moni95,dani03}.

For a given bulk density $\rho _{0}$, one can derive the beta-stability line 
$X=X^{\ast }(A,\rho _{0})$ by the condition 
\begin{equation}
\left. \frac{\partial E_{\mathrm{tot}}(\rho _{0},X)/A}{\partial X}
\right\vert _{A,\ X=X^{\ast }}=0.  \label{cond}
\end{equation}%
Near the beta-stability line, the total energy per particle (\ref{e1}) is
written up to the order $(X-X^{\ast })^{2}$ as 
\begin{equation}
E_{\mathrm{tot}}(\rho _{0},X)/A=E_{\mathrm{tot}}(\rho _{0},X^{\ast })/A+ %
\left[ b_{\mathrm{V,sym}}(\rho _{0})+b_{\mathrm{S,sym}}(\rho _{0})\
A^{-1/3}-\alpha _{C}(\rho _{0})A^{2/3}\right] (X-X^{\ast })^{2}.  \label{e2}
\end{equation}%
[Note that Eq. (\ref{e2}) is written for $A=\mathrm{constant}$.] The energy of
the ground state for a given value of mass number $A$ is obtained from the
additional equilibrium condition 
\begin{equation}
\left. \frac{\partial }{\partial \rho _{0}}E_{\mathrm{tot}}(\rho
_{0},X^{\ast })/A\right\vert _{A,\rho _{0}=\rho _{0,\mathrm{eq}}}=0,\quad
\label{var2}
\end{equation}%
where $\rho _{0,\mathrm{eq}}$ is$\ $the equilibrium bulk density.

The parameters $\Delta _{R}$ and $\Delta _{a}$ in the profile functions of
Eq. (\ref{prof1}) derive the neutron skin, $\Delta r_{np}$, and the neutron
excess, $N_{S}$, in the surface region of the nucleus (``neutron coating").
Substituting Eqs. (\ref{prof1}) and (\ref{fr}) into the conservation
particle condition and using \ the leptodermous expansion, we obtain for the
neutron excess $N-Z$ the following expression 
\begin{equation}
N-Z\approx N_{V}+N_{S},  \label{NZ1}
\end{equation}%
where%
\begin{equation}
N_{V}\approx \frac{4\pi }{3}R^{3}\left( 1+3\kappa _{0}(\eta )\frac{a}{R}%
+6\kappa _{1}(\eta )\left( \frac{a}{R}\right) ^{2}\right) \rho _{1},
\label{NZ2}
\end{equation}%
\begin{eqnarray}
N_{S} &\approx& 4\pi R^{2}\left[ \Delta _{R}\left( 1+2\kappa _{0}(\eta )
\frac{a}{R}+2\kappa_{1}(\eta )\left( \frac{a}{R}\right) ^{2}\right)\right. \nonumber \\
& &+ \left.\Delta _{a}\left( \kappa _{0}(\eta )+4\kappa _{1}(\eta )\frac{a}{R}
+3\kappa _{2}(\eta )\left( \frac{a}{R}\right) ^{2}\right) \right] \frac{\rho
_{0}}{2},
\end{eqnarray}
where $\kappa _{j}(\eta )$ are the generalized Fermi integrals derived in
Ref. \cite{kosa08} 
\begin{equation}
\kappa _{j}(\eta )=\int_{0}^{\infty }{dx\,x^{j}\left[ (1+e^{x})^{-\eta
}-(-1)^{j}\left( 1-(1+e^{-x})^{-\eta }\right) \right] }\ .  \label{ki}
\end{equation}%
The first term $N_{V}\sim R^{3}$ on the right hand side of Eq. (\ref{NZ1})
is caused by the redistribution of the neutron excess within the nuclear
volume while the second one $N_{S}\sim R^{2}$ is the neutron coating.

Note that the variational conditions of Eq. (\ref{var1}) leads to an
additional dependence of the variational parameters $\rho _{0}$, $\rho _{1}$%
, $R$, $a$, $\Delta _{R}$, $\Delta _{a}$ and $\eta $ on the external
parameters $A$ and $X$. The values of $\Delta _{R}$ and $\Delta _{a}$
depends slightly on the Skyrme
force parametrization. In the case of the \textrm{SkM} forces we have
evaluated the dependence of $\Delta _{R}$ and $\Delta _{a}$ on $X$ for $%
A=120 $ numerically and fitted it by the following formula 
\begin{equation}
\Delta _{R}(X)\approx 1.34\ X+0.07\ X^{2}\text{ }\mathrm{fm},\quad \Delta
_{a}(X)\approx 0.36\ X+\ 0.53\ X^{2}\text{ }\mathrm{fm}\text{ }.
\label{delta1}
\end{equation}

In general, the change of the radius $R$ of the nucleon distribution with
the nucleon number $A$ is caused by two factors. There is a simple
geometrical change of $R$ because of $R\propto A^{1/3}$. An additional
change can occur due to the polarization effect (the bulk density
distortion) with moving away the beta-stability line. In particular, the
size of the neutron skin is sensitive to the symmetry and Coulomb energies.
To see that we expand the total energy $E_{\mathrm{tot}}(\rho _{0},X)/A$
around the saturation density $\rho _{0,\mathrm{eq}}$. By keeping only terms
quadratic in $\delta \rho _{0}=\rho _{0}-\rho _{0,\mathrm{eq}}$ we rewrite
equilibrium Eq. (\ref{e2}) as \cite{kolu10} 
\begin{equation}
E_{\mathrm{tot}}(\rho _{0},X)/A=E_{\mathrm{tot}}(\rho _{0,\mathrm{eq}%
},X^{\ast })/A+\frac{K_{A}}{18\rho _{0,\mathrm{eq}}^{2}}(\rho _{0}-\rho _{0,%
\mathrm{eq}})^{2}+\frac{P_{A,\mathrm{sym}}}{\rho _{0,\mathrm{eq}}^{2}}%
(X-X^{\ast })^{2}(\rho _{0}-\rho _{0,\mathrm{eq}}),  \label{e3}
\end{equation}%
where $K_{A}$ is the incompressibility of finite nucleus 
\begin{equation}
K_{A}=9\left. \rho _{0,\mathrm{eq}}^{2}\frac{\partial ^{2}E_{\mathrm{tot}%
}(\rho _{0},X^{\ast })/A}{\partial \rho _{0}^{2}}\right\vert _{A,\rho
_{0}=\rho _{0,\mathrm{eq}}}  \label{ka}
\end{equation}%
and $P_{A,\mathrm{sym}}$ is the partial pressure related to the symmetry and
Coulomb energies%
\begin{equation}
P_{A,\mathrm{sym}}=\rho _{\rho _{0,\mathrm{eq}}}^{2}\left. \frac{\partial }{%
\partial \rho _{0}}\left[ b_{\mathrm{V,sym}}(\rho _{0})+b_{\mathrm{S,sym}%
}(\rho _{0})\ A^{-1/3}-\alpha _{C}(\rho _{0})A^{2/3}\right] \right\vert
_{A,\rho _{0}=\rho _{0,\mathrm{eq}}}.  \label{pa}
\end{equation}%
As seen from Eq. (\ref{e3}), a deviation from the beta-stability line ($%
X\neq X^{\ast }$) implies the change of the bulk density $\rho _{0}$. The
corresponding change of $\rho _{0}$ is dependent on the incompressibility $%
K_{A}$ and the partial pressure $P_{A,\mathrm{sym}}$. For an arbitrary fixed
value of $X$, the equilibrium density $\rho _{0,X}$ is derived by the
condition%
\begin{equation}
\left. \frac{\partial }{\partial \rho _{0}}E_{\mathrm{tot}}(\rho
_{0},X)/A\right\vert _{A,\rho _{0}=\rho _{0,X}}=0.  \label{eqq1}
\end{equation}%
Using Eqs. (\ref{e3}) and (\ref{eqq1}), we obtain the expression for the
shift of the bulk density (polarization effect) in the neutron rich nuclei 
\begin{equation}
\rho _{0,X}=\rho _{0,\mathrm{eq}}-9\frac{P_{A,\mathrm{sym}}}{K_{A}}%
(X-X^{\ast })^{2}.  \label{rho1}
\end{equation}

\begin{figure}[tbp]
\begin{center}
\includegraphics*[scale=0.6,clip]{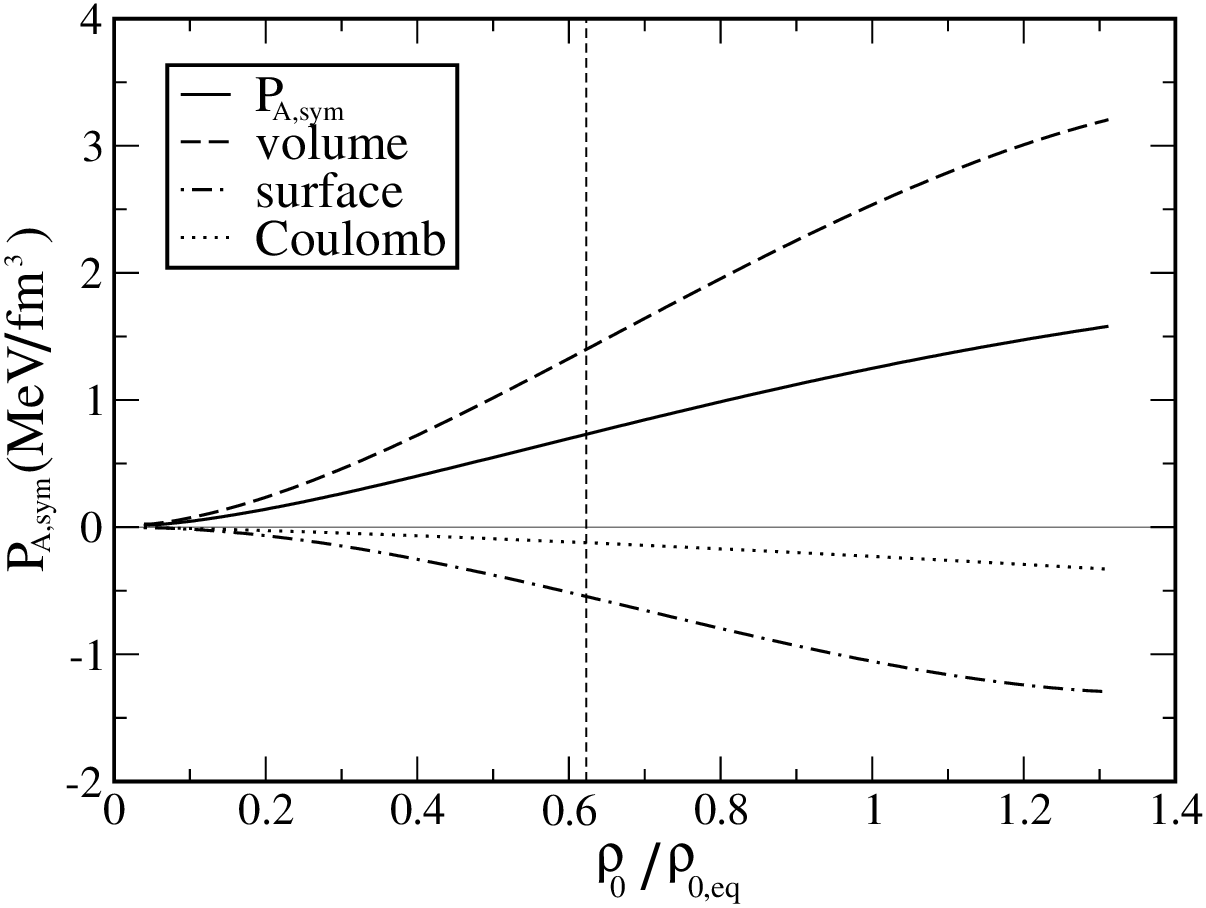}
\end{center}
\caption{The partial pressure $P_{A,\mathrm{sym}}$ for the nucleus $^{120}$%
\textrm{Sn} calculated for the SkM parametrization of the Skyrme force. The
equilibrium bulk density $\protect\rho _{0,eq}\simeq 0.153$ fm$^{-3}$. The
dashed vertical line is the spinodal instability border.}
\label{fig1}
\end{figure}

In Fig. \ref{fig1} we have plotted the partial pressure $P_{A,%
\mathrm{sym}}(\rho _{0})$ versus the bulk density $\rho _{0}$ (partial
equation of state) for the nucleus $^{120}Sn$. The partial contributions to $%
P_{A,\mathrm{sym}}(\rho _{0})$ \ from the symmetry volume $\sim $\textrm{\ }$%
\partial b_{\mathrm{V,sym}}(\rho _{0})/\partial \rho _{0}$, the symmetry
surface $\ \sim $\textrm{\ }$\partial b_{\mathrm{S,sym}}(\rho _{0})/\partial
\rho _{0}$, and the Coulomb $\ \sim \partial \alpha _{C}(\rho _{0})/\partial
\rho _{0}$ terms\ are also plotted in Fig. \ref{fig1}. The dashed
vertical line shows the position $\rho _{0}/\rho _{0,\mathrm{eq}}=0.62$ of
the spinodal instability border where $K_{A}=0$. On the left side of this
line the nucleus is unstable with respect to the bulk density variations. 

As seen from Fig. \ref{fig1}, the equilibrium partial pressure $P_{A,%
\mathrm{sym}}(\rho _{0,\mathrm{eq}})$ is positive and thereby $%
\rho_{0,X}<\rho _{0,\mathrm{eq}}$; see also Refs. \cite{oyta98,oyii03}. We
point out that in general the sign of the equilibrium partial pressure $P_{A,%
\mathrm{sym}}(\rho _{0,\mathrm{eq}})$ depends on the Skyrme force
parametrization and this fact can be used to fit the Skyrme forces \cite%
{brow01}.

\section{Radii of nucleon distributions and neutron skin}

As above noted, the bulk density $\rho _{0,X}$ is smaller for neutron-rich
nuclei; more neutrons should be pushed off to enrich the skin providing a
polarization effect. The nuclear \textrm{rms} radius 
\begin{equation}
\sqrt{\left\langle r^{2}\right\rangle }=\sqrt{\int {d\mathbf{r}\,r^{2}\,\rho
_{+}(r)/}\int {d\mathbf{r}\,\,\rho _{+}(r)}}.  \label{rms2}
\end{equation}
does not necessarily obey then the saturation condition having that $\sqrt{%
\left\langle r^{2}\right\rangle }$ is nonproportional to $A^{1/3}$. As a
consequence, the nuclei with significant excess of neutrons exhibit neutron
coating, i.e., are characterized by larger radii for the neutron than for
proton distributions. The interest in the neutron coating was recently raised
because of expectations that an analysis of the neutron coating could permit 
extrapolating the nuclear properties to neutron matter \cite{oyta98}. From the
point of view of study of the \textrm{EOS}, the coating size could provide
information on the derivative of the symmetry energy with respect to the
particle density \cite{oyta98,tybr01,furn02}. In general, the neutron coating 
$N_{S}$ of Eq. (\ref{NZ2}) can indicate the possibility of a giant neutron halo
which grows with moving away from the beta stability line \cite{meto02}. 
\begin{figure}[tbp]
\begin{center}
\includegraphics*[scale=0.6,clip]{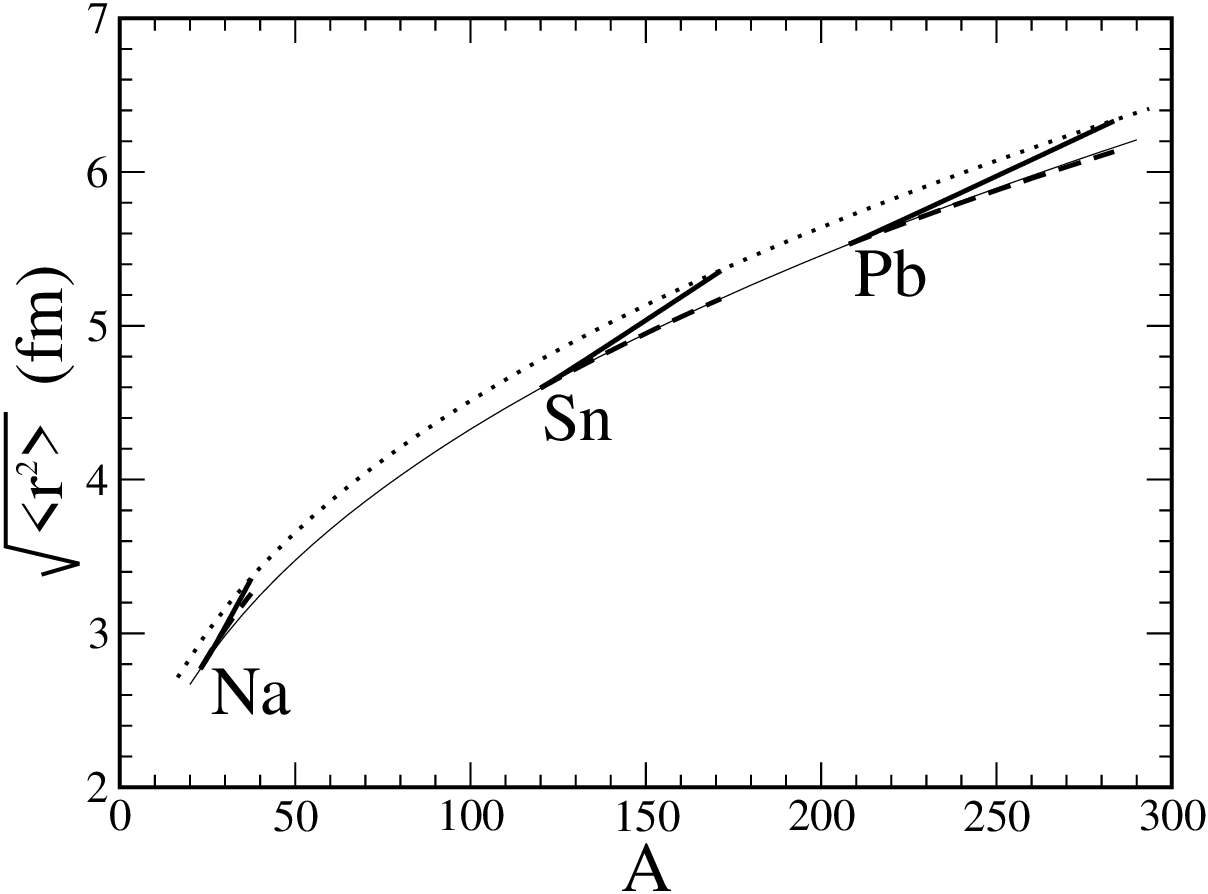}
\end{center}
\caption{The \textrm{rms} radius of nuclei near the beta-stability line. The
thin solid line is for the beta-stability line, the thick solid line is
beyond the beta-stability line for three nuclei, and the dotted curve is for
the neutron drip line. The dashed lines are the \textrm{rms} radii
calculated with the step distribution (\protect\ref{sdf}). The calculations
have been performed for the SkM parametrization of the Skyrme force.}
\label{fig2}
\end{figure}

In Fig. \ref{fig2} we have plotted (see the thick solid lines) the
nuclear \textrm{rms} radii $\sqrt{\left\langle r^{2}\right\rangle }$
obtained from Eq. (\ref{rms2}) for three nuclei. The results of Fig.
\ref{fig2} are only slightly sensitive to a small variation of the diffuse
layer and we have here assumed that $\Delta _{a}=0$. The thin solid line of %
Fig. \ref{fig2} represents the rms radius $\sim A^{1/3}$ along the
beta-stability line $X=X^{\ast }(A)$ which is parametrized by\ $X^{\ast
}(A)=0.17A^{2/3}\left/ \left( 26.5-25.6A^{-1/3}+0.17A^{2/3}\right) \right. $ 
\cite{kosa09}. The deviation of the rms radii (thick solid lines) from the
saturation behavior $\sim A^{1/3}$ (thin solid line) can not be related
directly to the appearance of the giant neutron halo at the approach to
the drip line (dotted curve) because we have here assumed $\Delta _{a}=0$.
As above noted, there are two sources for the change of the radius of
nucleon distribution with the nucleon number $A$. The first one is due to a
simple geometrical reason and the second one is because of the polarization
effect; see Eq. (\ref{rho1}).\textbf{\ }To extract a simple geometrical
change of the \textrm{rms} radius $\sqrt{\left\langle r^{2}\right\rangle }$
we will perform the calculations of $\sqrt{\left\langle r^{2}\right\rangle }$
with a step nucleon distribution 
\begin{equation}
\rho (r)=\rho _{0}\Theta (r-R).  \label{sdf}
\end{equation}%
Then the geometric \textrm{rms} radius calculated with the step function is
given by 
\begin{equation*}
\left. \sqrt{\left\langle r^{2}\right\rangle }\right\vert _{\mathrm{geom}}=%
\sqrt{\frac{3}{5}}R.
\end{equation*}%
We will normalize the ``geometrical" \textrm{rms} radius to the one $\left. 
\sqrt{\left\langle r^{2}\right\rangle ^{\ast }}\right\vert _{\mathrm{geom}}=%
\sqrt{3/5}R^{\ast }$, where $R^{\ast }$ is the nuclear radius on the
beta-stability line which obeys the saturation behavior $R^{\ast
}=r_{0}A^{1/3}$. Finally we obtain 
\begin{equation}
\left. \sqrt{\left\langle r^{2}\right\rangle }\right\vert _{\mathrm{geom}}=%
\sqrt{\frac{3}{5}}R^{\ast }\left( \frac{1-X^{\ast }}{1-X}\right) ^{1/3}.
\label{nrmsg}
\end{equation}%
The results of calculations\ by use of Eq. (\ref{nrmsg}) are shown in %
Fig. \ref{fig2} with the dashed lines. As one can see these results
for the nuclei $^{23}$\textrm{Na}, $^{120}$\textrm{Sn}, and $^{208}$\textrm{Pb%
} are very close to the ones on the beta-stability line (thin solid line).
The difference between the dashed lines and the thick solid ones represents
the magnitude of the polarization effect given by Eq. (\ref{rho1}). Thus, we
can conclude that the deviation of $\sqrt{\left\langle r^{2}\right\rangle }$
from the saturation behavior $\sim A^{1/3}$ in the regions of medium and
heavy nuclei is caused by the polarization effect which perturbs the
distribution of the neutron excess. 
\begin{figure}[tbp]
\begin{center}
\includegraphics*[scale=0.6,clip]{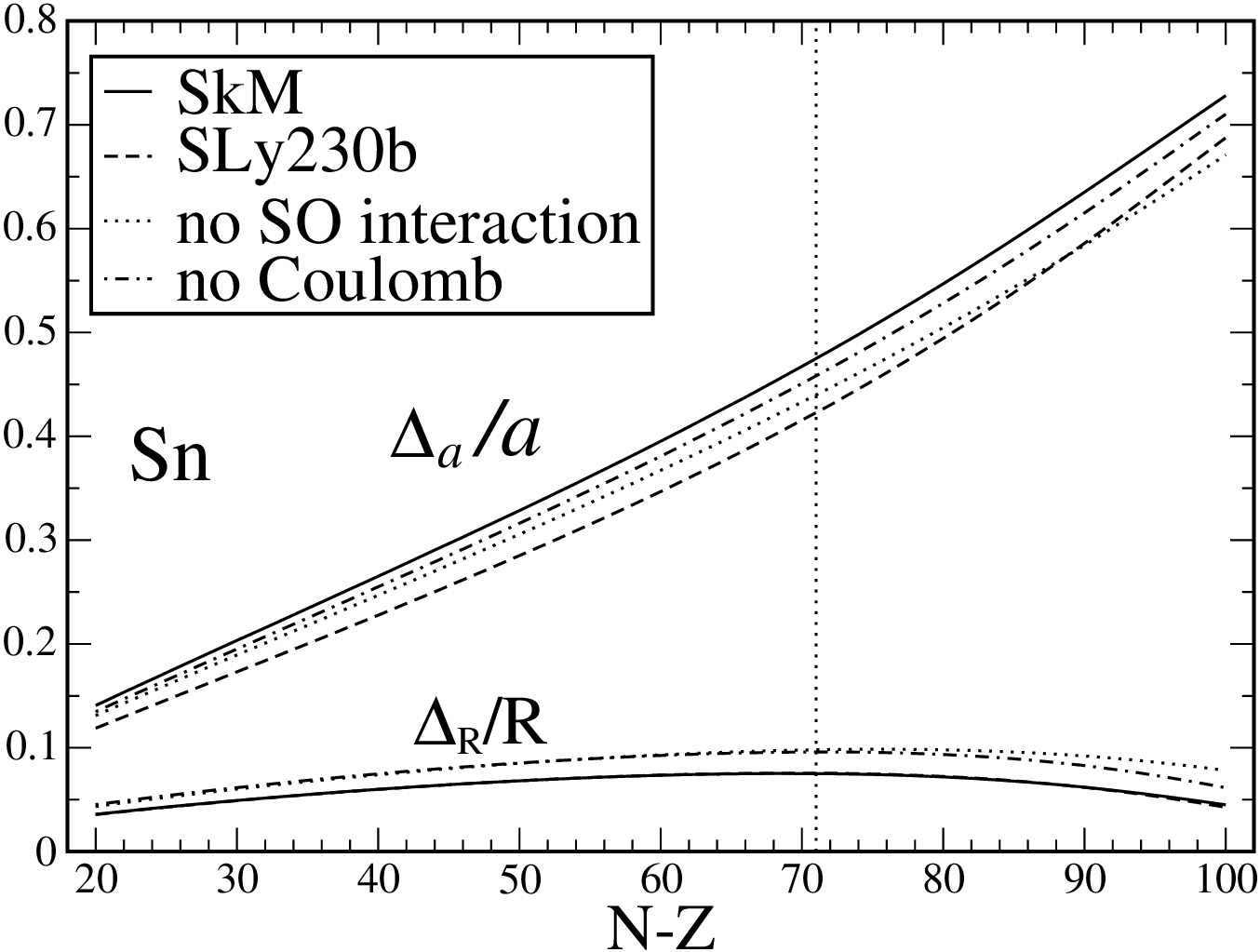}
\end{center}
\caption{Dependence of the relative shift of the neutron and proton radii 
$\Delta_{R}/R=(R_{n}-R_{p})/R$ and the diffuseness parameters $\Delta_{a}/a=(a_{n}-a_{p})/a$
versus the neutron excess $N-Z$ for the isotopes
with $Z=50$. The dotted line indicates no spin-orbit interaction and the dash-dotted
line no Coulomb interaction for the \textrm{SkM} parametrization. The
dashed line is for the \textrm{SLy230b} parametrization. The vertical dotted
line is for the neutron drip line.}
\label{fig3}
\end{figure}

To check the origin of the polarization effect, we have plotted in 
Fig. \ref{fig3} the relative shift of the neutron and proton radii $%
\Delta _{R}/R$ and the diffuseness parameters $\Delta _{a}/a$ versus the
neutron excess $N-Z$ for the fixed $Z=50$. Note that the direct use of the
variational procedure (\ref{var1}) with the modified profile functions $\rho
_{+}(r)$\ and $\rho _{-}(r)$\ of Eq. (\ref{prof1}) is badly converged with
respect to the variations of the parameter $\Delta _{a}$. To overcome this
difficulty in Fig. \ref{fig3}, we have performed the variational
calculations by use of the basic trial functions Eq. (\ref{rhoq}). One can
see from Fig. \ref{fig3} that the parameter $\Delta
_{a}/a=(a_{n}-a_{p})/a$, i.e., the parameter of the shape
distribution, is appreciably growing with the growing of the neutron coating
whereas the skin parameter $\Delta _{R}/R$ is only slightly sensitive to the
increase of $N-Z$. We point out also that both shift parameters $\Delta _{R}$
and $\Delta _{a}$ are only slightly sensitive to the semiclassical gradient
parameter $\beta $ in Eq. (\ref{ekin2}). 
\begin{figure}[tbp]
\begin{center}
\includegraphics*[scale=0.6,clip]{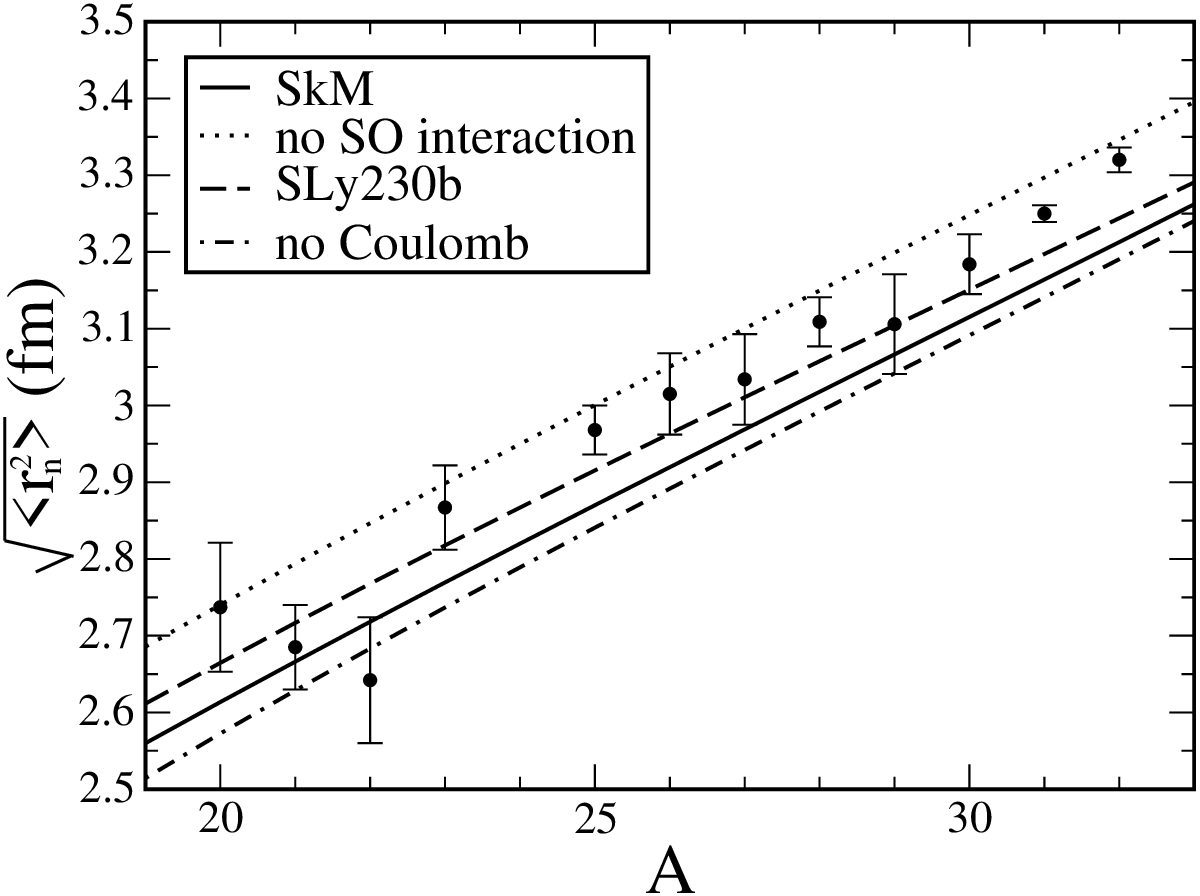}
\end{center}
\caption{The \textrm{rms} radius of neutron distribution in \textrm{Na}
isotopes for the \textrm{SkM} parametrization (solid line). The dotted line
indicates no spin-orbit interaction and the dash-dotted line no Coulomb interaction
for the \textrm{SkM} parametrization. The dashed line is for the \textrm{SLy230b} 
parametrization.}
\label{fig4}
\end{figure}

\begin{figure}[tbp]
\begin{center}
\includegraphics*[scale=0.6,clip]{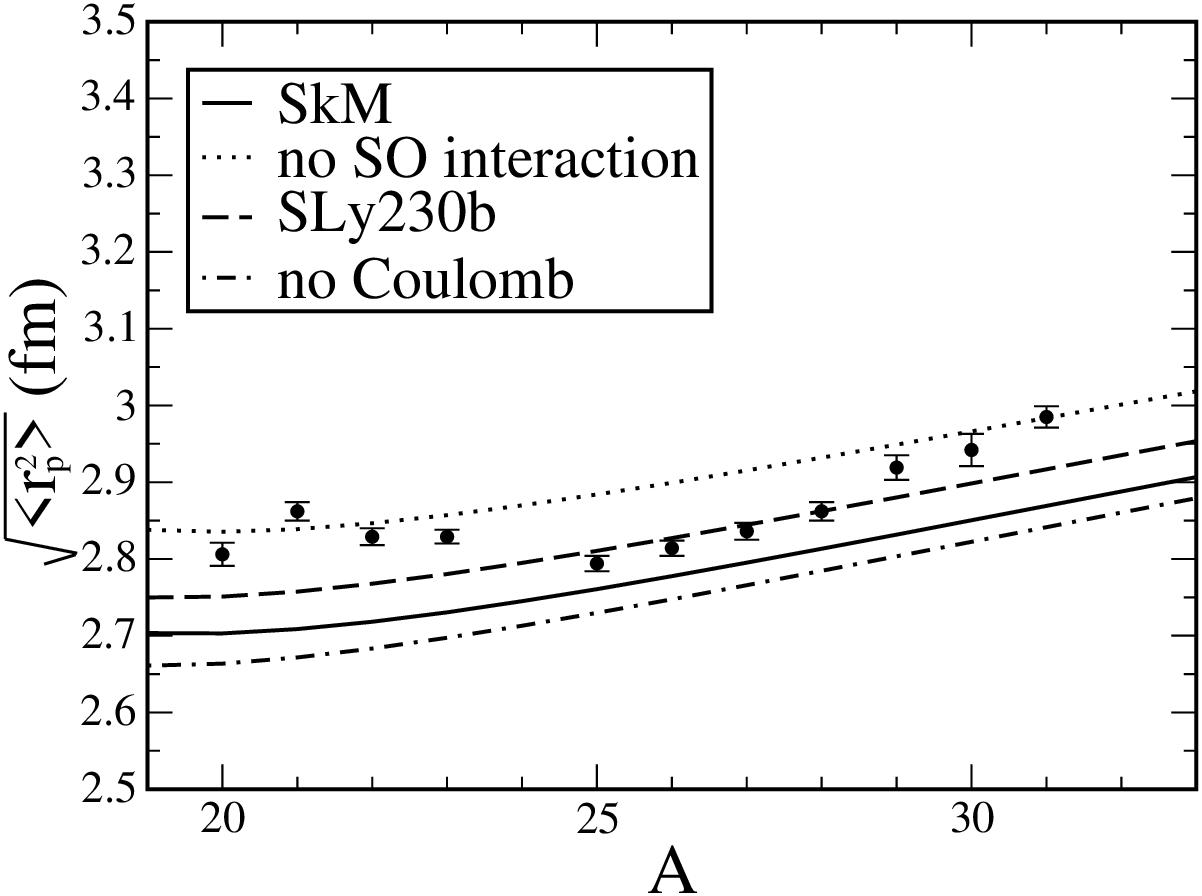}
\end{center}
\caption{The same as in Fig. 4 but for proton distribution.}
\label{fig5}
\end{figure}

The sensitivity of the rms radii of the nucleon distribution 
\begin{equation}
\sqrt{\left\langle r_{q}^{2}\right\rangle }=\sqrt{\int {d\mathbf{r}%
\,r^{2}\,\rho _{q}(r)/}\int {d\mathbf{r}\,\,\rho _{q}(r)}}.  \label{rms}
\end{equation}%
to the structure of the interparticle interaction is shown in 
Figs. \ref{fig4} and \ref{fig5} for \textrm{Na} isotopes. We can see
from Fig. \ref{fig4} that the ETFA results for $\sqrt{\left\langle
r_{n}^{2}\right\rangle }$ agree quite well with the experimental data from 
\cite{suge95}. The sensitivity of the calculation of
 $\sqrt{\left\langle r_{n}^{2}\right\rangle }$ to the choice of the Skyrme
forces for two parametrizations, \textrm{SkM} and \textrm{SLy230b}, can also 
be seen. Such kind of sensitivity can be used to fit the Skyrme force parameters. 
The two additional lines in Fig. \ref{fig4} show the influence of the
spin-orbit and Coulomb interactions on $\sqrt{\left\langle
r_{n}^{2}\right\rangle }$. As was mentioned above, the spin-orbit
interaction leads to the deeper potential in the surface region and
therefore a nuclear core attracts external coating neutrons. This effect
reduces the value of the polarization effect. The Coulomb interaction acts
in the opposite direction. The last is because the Coulomb interaction
increases the mean distance between nucleons and thereby the \textrm{rms}
radius.

The analogous results for the charge \textrm{rms} radius $\sqrt{\left\langle
r_{p}^{2}\right\rangle }$ of \textrm{Na} isotopes are shown in Fig.
\ref{fig5}. 
A small increase of \textrm{rms} radius $\sqrt{\left\langle
r_{p}^{2}\right\rangle }$ of the proton distribution with an increase of the
neutron number is caused by the neutron-proton attraction. Note that the
experimental data for proton \textrm{rms} radius $\sqrt{\left\langle
r_{p}^{2}\right\rangle }$ in Fig. \ref{fig5} manifests the
non-monotonic behavior which is due to the shell effects. The influence of
the spin-orbit and Coulomb interactions on the charge radius is the same as
in the previously observed case for the \textrm{rms} neutron radius. Note also
that such kind of behavior of $\sqrt{\left\langle r_{p}^{2}\right\rangle }$
is correlated with the $A$ dependence of the nuclear Coulomb radius $R_{C}$;
see Ref. \cite{kolu10}.

As mentioned in \textrm{Sect. 1}, the value of isotopic shift of
radii $\Delta r_{np}$\ can be caused by both the skin effect and the halo
effect in the neutron and proton distributions. To separate these effects,
we will represent the value of $\Delta r_{np}$\ as%
\begin{equation}
\Delta r_{np}=\Delta r_{np,R}+\Delta r_{np,a},  \label{drho1}
\end{equation}%
where $\Delta r_{np,R}$ and $\Delta r_{np,a}$\ are caused by the different
radii (skin effect) and the different diffuseness (halo effect) of 
the neutron and proton distributions, respectively. The corresponding
values are given by 
\begin{equation}
\Delta r_{np,R}\approx \sqrt{\frac{3}{5}}\left\{ 1+\frac{7}{2}\left[ \kappa
_{0}^{2}(\eta )-2\kappa _{1}(\eta )\right] \left( \frac{a}{R}\right)
^{2}\right\} \Delta _{R},  \label{drhoR}
\end{equation}%
and%
\begin{equation}
\Delta r_{np,a}\approx \sqrt{\frac{3}{5}}\left\{ \kappa _{0}(\eta )-7\left[
\kappa _{0}^{2}(\eta )-2\kappa _{1}(\eta )\right] \frac{a}{R}\right\} \Delta
_{a},  \label{drhoa}
\end{equation}

\begin{figure}[tbp]
\begin{center}
\includegraphics*[scale=0.6,clip]{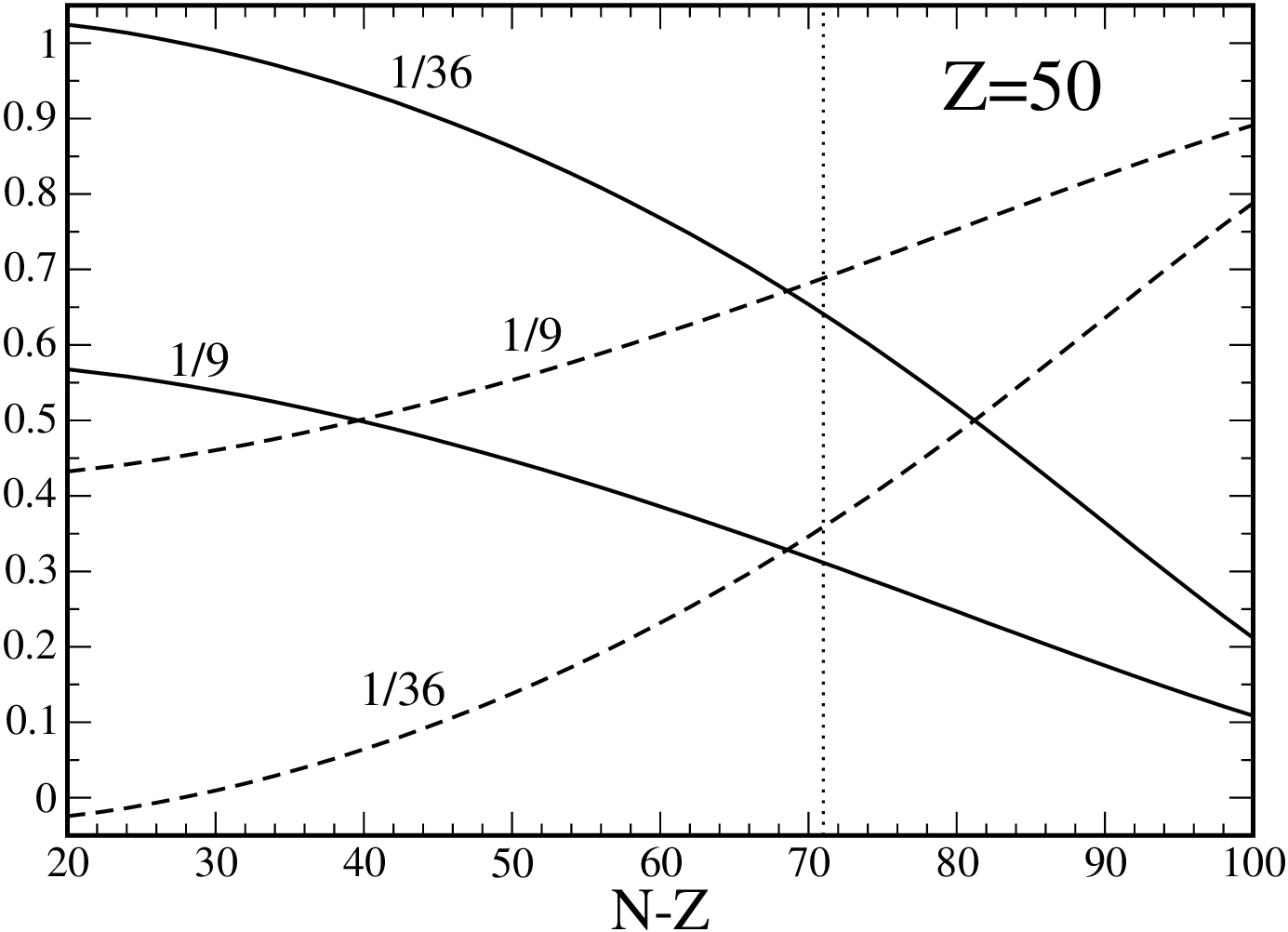}
\end{center}
\caption{The partial contribution to the isotopic shift of radii $\Delta
r_{np}$ from the skin effect, $\Delta r_{np,R}/\Delta r_{np}$ (solid lines),
and the halo effect, $\Delta r_{np,a}/\Delta r_{np}$\ (dashed lines), versus
the neutron excess in \textrm{Sn} isotopes. The calculations have been
performed for two values of the gradient parameter $\protect\beta =1/36$ and 
$\protect\beta =1/9$ indicated near the corresponding lines. The vertical
dotted line is for the neutron drip line.}
\label{fig6}
\end{figure}

In Fig. \ref{fig6} we have plotted the values of $\Delta r_{np,R}$\
and $\Delta r_{np,a}$\ versus the neutron excess for \textrm{Sn} isotopes. 

As can be seen in Fig. \ref{fig6}, the relative contribution of the
shape (halo) effect, i.e., $\Delta r_{np,a}$, to the isotopic shift of radii $%
\Delta r_{np}$\ depends strongly on the parameter $\beta$ of the diffuse tail
in the nucleon density distribution; see Eq. (\ref{rho1x}). For the
semiclassical value of $\beta =1/36$, the halo effect is quite small near
the beta stability line and can play an appreciable role close to the drip
line only. The situation is significantly different in the case of the
phenomenological value of $\beta =1/9$ where the contribution of $\Delta
r_{np,a}$, and thereby the halo effect, to $\Delta r_{np}$\ is more
appreciable.

\begin{figure}[tbp]
\begin{center}
\includegraphics*[scale=0.6,clip]{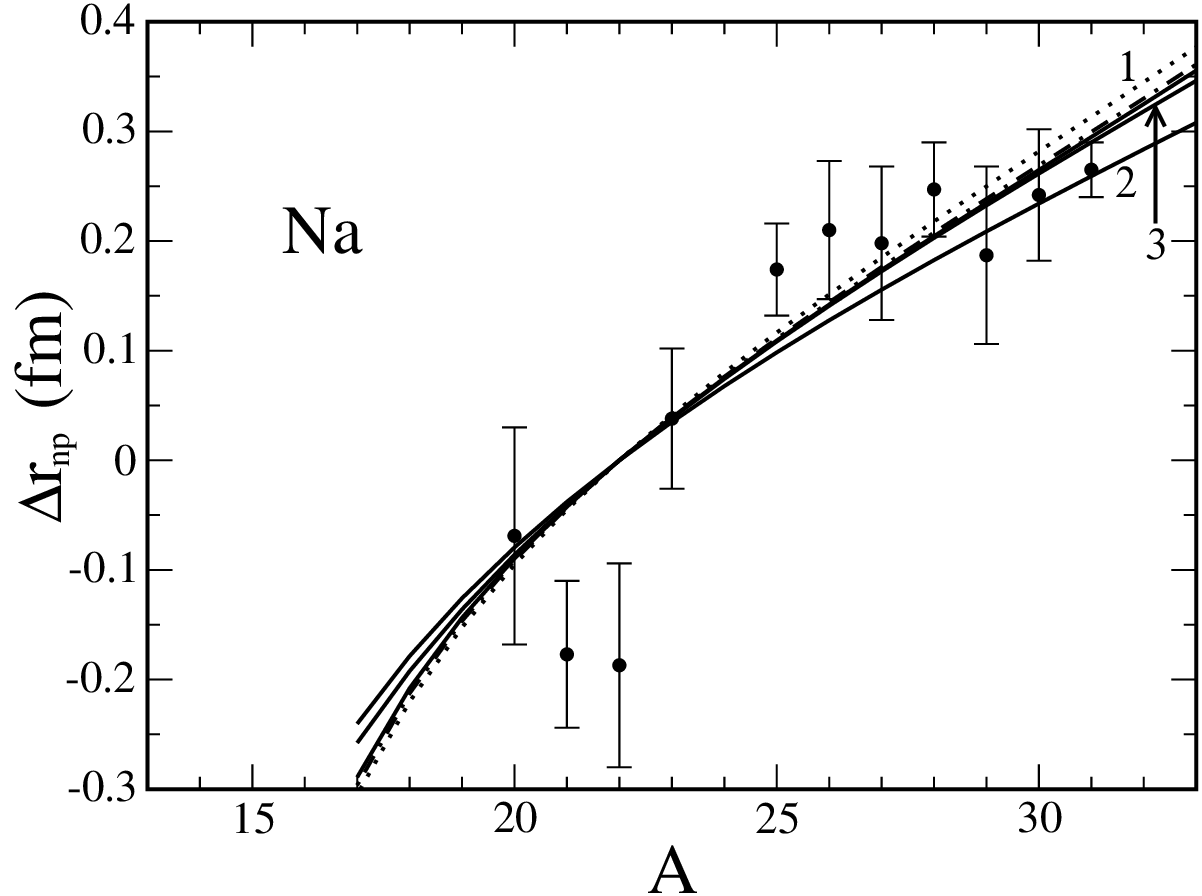}
\end{center}
\caption{Isovector shift of nuclear rms radius $\Delta r_{np}=\protect\sqrt{%
\left\langle r_{n}^{2}\right\rangle }-\protect\sqrt{\left\langle
r_{p}^{2}\right\rangle }$ in \textrm{Na} isotopes for the \textrm{SkM}
parametrization. Solid line 1 was obtained by use of trial functions Eq. 
(\protect\ref{rhoq}), i.e., $\Delta _{a}\neq 0$, and $\protect\beta =1/36$.
The dotted line is the same but without spin-orbit interaction and the
dash-dotted line is without Coulomb interaction. The solid line 2 is for $%
\Delta _{a}=0$ and $\protect\beta =1/36$; the solid line 3 is for $\Delta
_{a}=0$ and $\protect\beta =1/9$.}
\label{fig7}
\end{figure}

\begin{figure}[tbp]
\begin{center}
\includegraphics*[scale=0.6,clip]{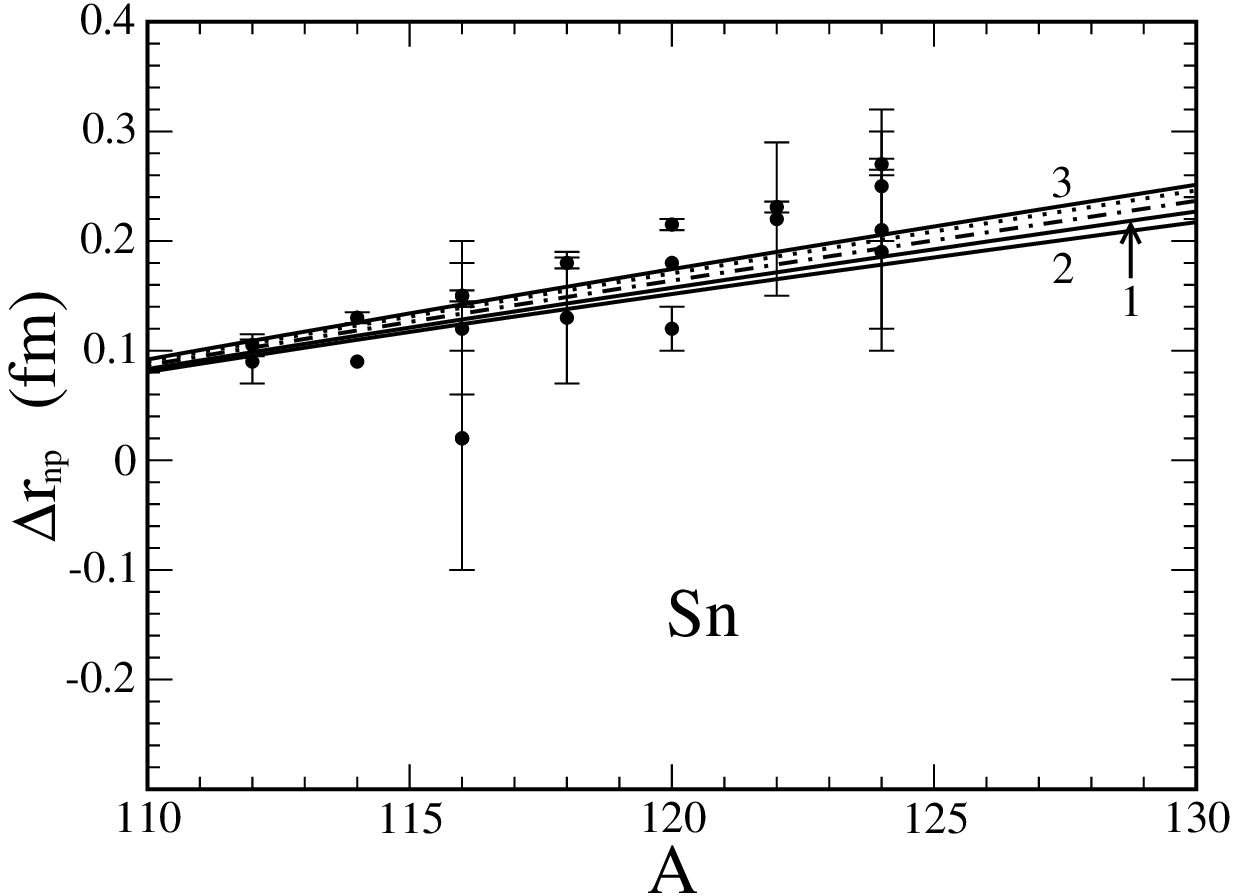}
\end{center}
\caption{The same as in Fig. \ref{fig7} but for \textrm{Sn}
isotopes.}
\label{fig8}
\end{figure}

\begin{figure}[tbp]
\begin{center}
\includegraphics*[scale=0.6,clip]{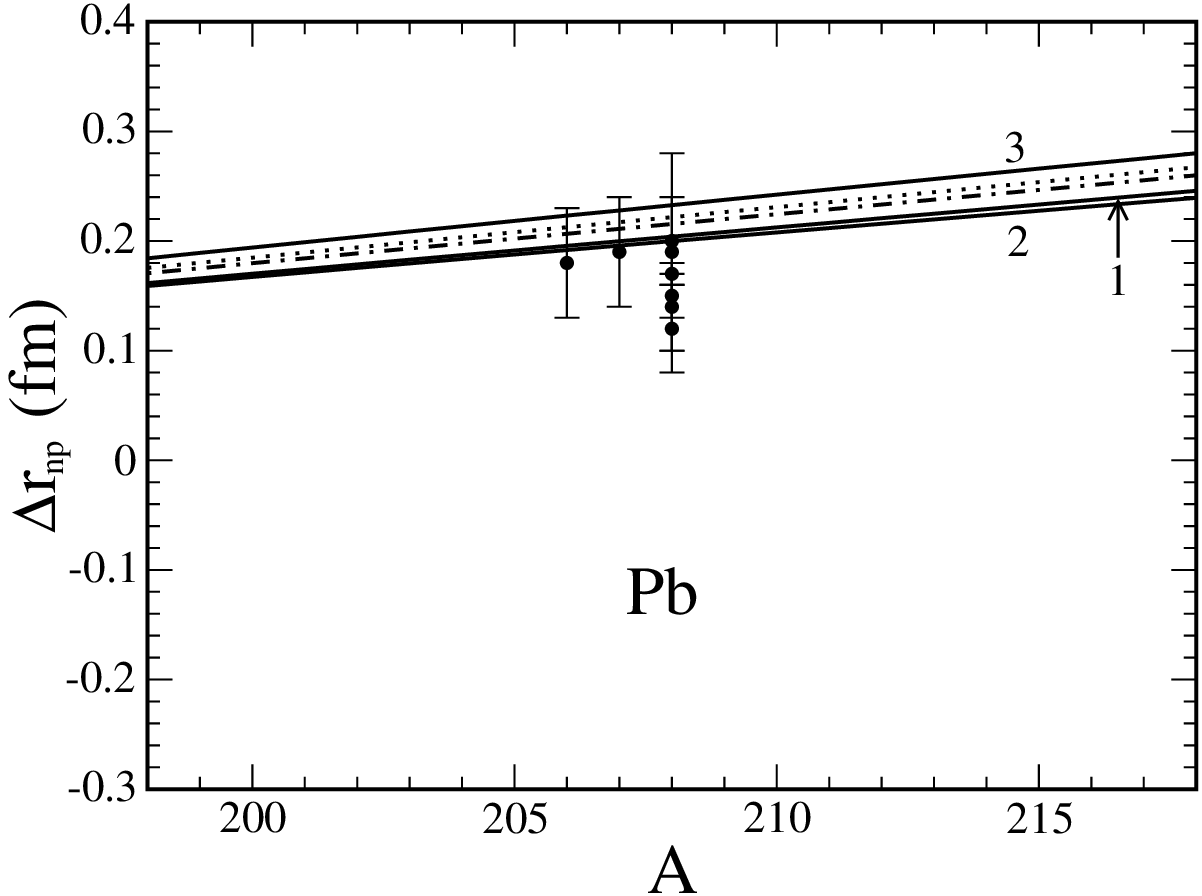}
\end{center}
\caption{The same as in Fig. \ref{fig7} but for \textrm{Pb} isotopes.}
\label{fig9}
\end{figure}

The $A$ dependence of the size of the neutron coating $\Delta r_{np}=\sqrt{%
\left\langle r_{n}^{2}\right\rangle }-\sqrt{\left\langle
r_{p}^{2}\right\rangle }$ is illustrated in Figs. \ref{fig7}, \ref{fig8}, and 
\ref{fig9} for \textrm{Na}, \textrm{Sn}, and \textrm{Pb} isotopes, respectively. 
The numerical results have been obtained from Eq. (\ref{rms}) by use of the
basic trial functions Eq. (\ref{rhoq}). The experimental data have been
taken from Refs. \cite{suge95,ray79,sthi94,kaam02,clke03}. 

\bigskip

As can be seen from Figs. \ref{fig7}, \ref{fig8}, and \ref{fig9}, the
Coulomb interaction affects the isovector shift of nuclear radii weakly
but with growing of $A$ and $X$ this influence slightly increases. The last
is because the Coulomb interaction increases the distance between protons,
i.e., $\left\langle r_{p}^{2}\right\rangle $, and reduces thereby the
isovector shift. The spin-orbit interaction produces the same effect as the
Coulomb interaction but with stronger magnitude. As was mentioned above,
the spin-orbit interaction leads to a deeper potential near the surface
region and the nuclear core attracts the external neutrons decreasing the
diffuse layer of the neutron distribution. That reduces the isovector shift
of nuclear radii because of Eqs. (\ref{drho1}) and (\ref{drhoa}). The
spin-orbit effect on $\Delta r_{np}$ increases with $X$ because the increase
of $X$\ leads to the contribution to the density $\rho _{n}(r)$ of neutrons
with higher angular momentum.

The value of the halo effect in $\Delta r_{np}$ can be estimated from 
Figs. \ref{fig7}, \ref{fig8}, and \ref{fig9} by comparison of the
solid lines \textrm{1} and \textrm{2}. The curve \textrm{2} was obtained
neglecting the contribution from the isovector diffuseness distortion, i.e.,
for $\Delta _{a}=0$. We can see that the halo effect, which occurs because
of $\Delta _{a}\neq 0$, is quite small even for the lighter nucleus \textrm{%
Na}. As seen in Figs. \ref{fig7}, \ref{fig8}, and \ref{fig9}, the halo
effect leads to an increase of the isovector shift $\Delta r_{np}$ of
nuclear radii. The change of the slope of curves $\Delta r_{np}(A)$ due to
the halo effect can be also obtained analytically from Eqs. (\ref{drho1}), 
(\ref{drhoR}), and (\ref{drhoa}). Taking into account a typical dependency of
the values $\Delta _{R}(X)$ and $\Delta _{a}(X)$ on the asymmetry parameter $%
X$, see e.g. Eq. (\ref{delta1} ), and the fact that $\kappa _{0}(\eta )=-1$
and $\kappa _{1}(\eta )\simeq 1.65$ for $\eta =2$, one can see that the
slope of curves $\Delta r_{np}(A)$ grows with an increase of the diffuse
layer shift $\Delta_{a}$. Comparing the relative shifts of curves \textrm{1}
and \textrm{2} in Figs. \ref{fig7}, \ref{fig8}, and \ref{fig9}, one
can also conclude that the halo effect is reduced strongly for the heavier
nuclei. This is because of a general increase of the nuclear stiffness with
respect to the variation of the diffuse layer with growing $A$.

A growth of the parameter $\beta $ which is responsible for the diffuse tail
in the nucleon density distribution, see Eq. (\ref{rho1x}), increases the
slope of the curve $\Delta r_{np}(A)$; see lines \textrm{2} and \textrm{3}
in Figs. \ref{fig7}, \ref{fig8}, and \ref{fig9}. This is caused by the
fact that the diffuse layer of the neutrons exceeds the one for the protons
and the multiplication of the nucleon density ${\rho _{q}(r)}$ by a factor of 
$r^{2}$ in Eq. (\ref{rms}) leads to an emphasis of the peripheral region of
the particle density which is stronger for the neutron peripheral region
than for the proton one. Note also that the crossing point in Fig.
\ref{fig7} for isotopes of \textrm{Na} happens for $\Delta r_{np}=0$ where $%
N=Z$.

\section{Summary}

We have applied the direct variational method within the extended
Thomas-Fermi approximation with effective Skyrme-like forces to the
description of the radii of nucleon distributions. In our consideration, the
thin-skinned nucleon densities $\rho _{p}(\mathbf{r})$ and $\rho _{n}(%
\mathbf{r})$ are generated by the profile functions which are eliminated by
the requirement that the energy of the nucleus should be stationary with
respect to variations of these profiles. An advantage of the used direct
variational method is the possibility to derive the equation of state for
finite nuclei: dependence of the binding energy per particle or the pressure
on the bulk density $\rho _{0}$. We have evaluated the partial pressure $%
P_{A,\mathrm{sym}}$ which includes the contributions from the symmetry and
Coulomb energies. The pressure $P_{A,\mathrm{sym}}$ is positive driving off
the neutrons in neutron-rich nuclei to the skin.

Using the leptodermous properties of the profile nucleon densities $\rho
_{p}(\mathbf{r})$ and $\rho _{n}(\mathbf{r})$, we have established the
presence of the neutron coating $N_{S}$. The size of the neutron coating is
growing with moving away from the beta stability line. In Fig.
\ref{fig2} this fact is demonstrated as a deviation of the \textrm{rms} radius
of the nucleon distribution from the saturation behavior $\sim A^{1/3}$ in
the nuclei beyond the beta-stability line. Moreover, the neutron skin
develops by diffusing the neutron surface against the changeless proton
diffuseness and can be responsible for the giant neutron halo in
neutron-rich nuclei.

The average behavior of the nucleon distribution $\sqrt{\left\langle
r_{q}^{2}\right\rangle }$ and the size of the neutron skin is satisfactorily
described within the extended Thomas-Fermi approximation. The sensitivity of
the calculations of nuclear \textrm{rms} radii $\sqrt{\left\langle
r^{2}\right\rangle }$ to the choice of the force parametrization can be used
to a fit of Skyrme forces. We have pointed that the charge radii of proton
distributions show the shell oscillations with $A$ which are related to the
shell effects in the Coulomb energy. The charge radii are connected to the
isospin shift of neutron-proton chemical potentials $\Delta \lambda =\lambda
_{n}-\lambda _{p}$ for nuclei beyond the beta-stability line by fixed value
of the mass number $A$ \cite{kolu10}. It was shown that the isovector shift
of the nuclear radius $\Delta r_{np}$ is primarily linear dependent on the
asymmetry parameter $X$. The Coulomb and spin-orbit interactions do not
affect significantly the isovector shift of the nuclear radius.

We have established the influence of the polarization effect given by Eq. (%
\ref{rho1}) on the rms radius $\sqrt{\left\langle r_{n}^{2}\right\rangle }$
of the neutron distribution. This effect increases with the asymmetry
parameter $X$ and can be responsible for the appearance of the giant neutron
halo in the nuclei close to the drip line. We have also estimated the
relative contribution to the value of the isotopic shift of radii $\Delta
r_{np}$ obtained from both the skin effect and the halo effect. The halo
effect gives usually a minor contribution to the shift $\Delta r_{np}$ and
it can be comparable with the skin effect near the drip line only.


\begin{thebibliography}{99}
\bibitem{meto02} 
J. Meng, H. Toki, J.Y. Zeng, S.Q. Zhang and S.-G. Zhou,
Phys. Rev. C \textbf{65}, 041302 (2002).

\bibitem{midolanare00} 
S. Mizutori, J. Dobaczewski, G.A. Lalazissis, W. Nazarewicz and P.-G. Reinhard, 
Phys. Rev. C \textbf{61}, 044326 (2000).

\bibitem{trja01} 
A. Trzci\'{n}ska, J. Jastrz\'{e}bsky, P. Lubi\'{n}ski, F. J. Hartmann, R. Schmidt, 
T. von Egidy, and B. K\l{}os,
Phys. Rev. Lett. \textbf{87}, 082501 (2001).

\bibitem{krak04} 
A. Krasznahorskay, H. Akimune, A.M. van den Berg, \textit{et al}.,
Nucl. Phys. \textbf{A731}, 224 (2004).

\bibitem{waviroce10} 
M. Warda, X. Vi\~nas, X. Roca-Maza and M. Centelles,
Phys. Rev. C \textbf{81}, 054309 (2010).

\bibitem{ceroviwa10} 
M. Centelles, X. Roca-Maza, X. Vi\~nas and M. Warda,
Phys. Rev. C \textbf{82}, 054314 (2010).

\bibitem{niscnabepe11} 
N. Nikolov, N. Schunck, W. Nazarewicz, M. Bender and J. Pei, 
Phys. Rev. C \textbf{83}, 034305 (2011).

\bibitem{gansamo11} 
M.K. Gaidarov, A.N. Antonov, P. Sarriguren, and E. Moya de Guerra, 
Phys. Rev. C \textbf{84}, 034316 (2011).

\bibitem{hoko64} 
P. Hohenberg and W. Kohn, 
Phys. Rev. \textbf{136}, B 864 (1964).

\bibitem{kirz67} 
D.A. Kirzhnitz, 
\textit{Field Theoretical Methods in Many Body-Systems} (Pergamon, London, 1967).

\bibitem{brguha85} 
M. Brack, C. Guet and H.-B. H\aa kansson, 
Phys. Rep. \textbf{123}, 275 (1985).

\bibitem{book} 
V.M. Kolomietz, 
\textit{Local Density Approach for Atomic and Nuclear Physics} (Naukova Dumka, Kiev, 1990; in Russian).

\bibitem{kosa08} 
V.M. Kolomietz and A.I. Sanzhur, 
Eur. Phys. J. \textbf{A38}, 345 (2008).

\bibitem{mysw69} 
W.D. Myers and W.J. Swiatecki, 
Ann. Phys. (NY) \textbf{55}, 395 (1969).

\bibitem{mysw74} 
W.D. Myers and W.J. Swiatecki, 
Ann. Phys. (NY) \textbf{84}, 186 (1974).

\bibitem{moni95} 
P. M\"{o}ller, J.R. Nix, W.D. Myers and W.J. Swiatecki, 
At. Data Nucl. Data Tables \textbf{59}, 185 (1995).

\bibitem{dani03} 
P. Danielewicz, 
Nucl. Phys. \textbf{A} \textbf{727}, 233 (2003).

\bibitem{kolu10} 
V.M. Kolomietz, S.V. Lukyanov and A.I.Sanzhur, 
Nucl. Phys. At. Energy \textbf{11}, 335 (2010).

\bibitem{oyta98} 
K. Oyamatsu, I. Tanichata, S. Sugahara, K. Sumiyoshi and H. Toki, 
Nucl. Phys. \textbf{A} \textbf{634}, 3 (1998).

\bibitem{oyii03} 
K. Oyamatsu and K. Iida, 
Progr. Theor. Phys. \textbf{109}, 631 (2003).

\bibitem{brow01} 
B.A. Brown, 
Phys. Rev. Lett. \textbf{85}, 5296 (2000).

\bibitem{tybr01} 
S. Typel and B.A. Brown, 
Phys. Rev. C \textbf{64}, 027302 (2001).

\bibitem{furn02} 
R.J. Furnstahl, 
Nucl. Phys. \textbf{A} \textbf{706}, 85 (2002).

\bibitem{kosa09} 
V.M. Kolomietz and A.I. Sanzhur, Phys. Rev. C \textbf{81}, 024324 (2010).

\bibitem{suge95} 
T. Suzuki, H. Geissel, O. Bochkarev, \textit{et al}., 
Phys. Rev. Lett. \textbf{75}, 3241 (1995).


\bibitem{ray79} 
L. Ray, 
Phys. Rev. C \textbf{19} , 1855 (1979).

\bibitem{sthi94} 
V.E. Starodubsky, N.M. Hintz, 
Phys. Rev. C \textbf{49}, 2118 (1994).

\bibitem{kaam02} 
S. Karataglidis, K. Amos, B.A. Brown, P.K. Deb, 
Phys. Rev. C \textbf{65}, 044306 (2002).

\bibitem{clke03} 
B.C. Clark, L.J. Kerr, S. Hama, 
Phys. Rev. C \textbf{67}, 054605 (2003).
\end{thebibliography}
\end{document}